\documentclass[aps,pra,twocolumn,a4paper,groupedaddress,floatfix,10pt,longbibliography]{revtex4-1}

\usepackage{enumitem}
\usepackage{amsmath}
\usepackage{amssymb}
\usepackage{amsfonts}
\usepackage{braket}
\usepackage{dsfont}
\usepackage{MnSymbol} 
\newcommand{\imag}{\mathrm{i}}
\usepackage{color}

\usepackage{graphicx}
\usepackage[utf8]{inputenc}
\usepackage[export]{adjustbox}
\usepackage[caption=false]{subfig}
\usepackage{parskip} 

\begin{document}
\title{Minimum Hardware Requirements for Hybrid Quantum-Classical DMFT}

\author{B. Jaderberg$^{1}$, A. Agarwal$^{1}$, K. Leonhardt$^{1}$, M. Kiffner$^{1,2}$ and D. Jaksch$^{1,2}$}
\affiliation{$^1$Clarendon Laboratory, University of Oxford, Parks Road, Oxford OX1 3PU, United Kingdom} 
\affiliation{$^2$Centre for Quantum Technologies, National University of Singapore, 3 Science Drive 2, Singapore 117543}
\date{\today}
\begin{abstract}
 We numerically emulate noisy intermediate-scale quantum (NISQ) devices and determine the minimal hardware requirements for two-site hybrid quantum-classical dynamical mean-field theory (DMFT). We develop a circuit recompilation algorithm which significantly reduces the number of	quantum gates of the DMFT algorithm and find that the quantum-classical algorithm converges	if the two-qubit gate fidelities are larger than 99\%. The converged results agree with the exact solution within 10\%, and perfect agreement within noise-induced error margins can be obtained for	two-qubit gate fidelities exceeding 99.9\%. By comparison, the quantum-classical algorithm without	circuit recompilation requires a two-qubit gate fidelity of at least 99.999\% to achieve perfect agreement with the exact solution. We thus find quantum-classical DMFT calculations can be run on the next generation of NISQ devices if combined with the recompilation techniques developed in this work.
\end{abstract}

\maketitle

\section{Introduction} \label{sec:introduction}
Scalable, fault-tolerant quantum computers promise to solve problems that are 
intractable on classical computers such as the simulation of quantum systems~\cite{lloyd_1996} or factorising composite integers~\cite{shor_1997}. Ongoing efforts to build a quantum computer are currently in the noisy intermediate-scale quantum (NISQ) era, characterised by hardware with less than 100 qubits, large gate errors and no error correction~\cite{preskill_2018}. 

In general, NISQ devices are believed to be well suited to solving optimisation problems using hybrid quantum-classical algorithms~\cite{peruzzo_mcclean_shadbolt_yung_zhou_love_aspuru-guzik_obrien_2014,mcclean_romero_babbush_aspuru-guzik_2016, guerreschi_smelyanskiy_2017}. In these,  a cost function is encoded into  a quantum circuit with parameterised quantum logic gates, and a classical algorithm iteratively optimises these parameters to minimise or maximise the cost function. Variational quantum algorithms  have been successfully applied to a number of problems on existing NISQ devices. For example, quantum chemistry calculations were carried out on  superconducting~\cite{malley_babbush_kivlichan_romero_2016, kandala_mezzacapo_temme_takita_brink_chow_gambetta_2017, colless_ramasesh_dahlen_blok_kimchi-schwartz_mcclean_carter_dejong_siddiqi_2018} and  ionic~\cite{hempel_maier_romero_mcclean_2018} NISQ devices, and nuclear structure calculations were performed on quantum processors accessed via cloud servers~\cite{dumitrescu_mccaskey_hagen_jansen_2018}. Furthermore, nuclear magnetic resonance systems were used to demonstrate a hybrid quantum-classical approach to quantum optimal control~\cite{li_yang_peng_sun}.

The success of NISQ devices in solving small-scale electronic structure problems is substantiated by theoretical results showing that quantum computers can solve correlated electronic structure problems in polynomial time~\cite{lloyd_1996, aspuru-guzik_dutoi_love_head-gordon_2005}, e.g., via phase estimation algorithms~\cite{lanyon_whitfield_gillett_goggin_2010}. It is therefore natural to consider if other electronic structure methods could benefit from a quantum computational approach. For example, dynamical mean-field theory (DMFT)~\cite{georges_kotliar_krauth_rozenberg_1996} is a standard approach for simulating materials with strong electronic correlations, and proposals for hybrid quantum-classical DMFT algorithms have been put forward recently~\cite{kreula_clark_jaksch_2016,kreula_garcia-alvarez_lamata_clark_solano_jaksch_2016,bauer_wecker_millis_hastings_troyer_2016}. Experimental realisations have been achieved for the insulating phase~\cite{keen_maier_johnston_lougovski_2020} and in the case of an alternative approach to DMFT \cite{rungger_fitzpatrick_chen_alderete_apel_2019}, which uses the variational quantum eigensolver method to calculate ground and excited states of the system. However, to the best of our knowledge the precise hardware requirements for obtaining high-quality DMFT results on  a quantum computer are not known. 

Here we determine the hardware requirements of hybrid quantum-classical DMFT by numerically emulating NISQ devices via the Qiskit framework~\cite{Qiskit}. Specifically, we consider the two-site DMFT scheme in~\cite{kreula_garcia-alvarez_lamata_clark_solano_jaksch_2016}, which forms a basic building block of a scalable and digital quantum computing approach to DMFT. Our noise modelling takes into account finite  qubit lifetimes as well as gate and measurement errors. We find that the quantum-classical algorithm produces solutions that agree with the exact results within a few percent if the two-qubit gate fidelity exceeds 99.99\%. Increasing the two-qubit gate fidelities beyond  99.999\% allows one to achieve perfect agreement with the exact solution apart from noise-induced, residual errors. 

Furthermore, we show that these stringent error bounds can be substantially relaxed by applying recent results in quantum circuit recompilation \cite{jones_benjamin_2018, khatri_larose_poremba_cincio_sornborger_coles_2019,ostaszewski_mateusz_grant_benedetti_2019}, to significantly reduce the number of gates in the quantum DMFT circuit. In this way we find that two-qubit gate fidelities exceeding 99\% or 99.9\% are sufficient for quantum-classical DMFT calculations with 10\% error or perfect agreement to the exact results respectively. It follows that these calculations could therefore be run on next-generation NISQ devices. 

This paper is organised as follows. In Sec.~\ref{sec:model} we introduce our model for running hybrid quantum-classical DMFT algorithms on NISQ devices. Our results for the minimal hardware requirements of two-site DMFT are presented in Sec.~\ref{sec:results}. We first consider the hardware requirements of the full scheme   in Sec.~\ref{subsec:fidelity_requirements_original}, and then show how techniques for reducing the circuit depth can dramatically reduce these requirements in Sec.~\ref{subsec:isl}. In Sec.~\ref{sec:conclusion} we review our findings  and look at the possible future of running hybrid quantum-classical DMFT on NISQ hardware.
	
\section{Model}\label{sec:model}
In this section we present the model for determining the minimal hardware 
requirements of hybrid quantum-classical DMFT. For this we give a 
very brief introduction to Hamiltonian-based DMFT in 
Sec.~\ref{subsec:siam_hamiltonian}. In 
Sec.~\ref{subsec:quantum_circuits}, we explain the individual steps that make up hybrid quantum-classical DMFT and describe the required  quantum 
circuits. Finally, in Sec.~\ref{subsec:noise_models}, we detail the construction of the noise model used to simulate errors like those seen in real quantum hardware.
\subsection{SIAM Hamiltonian}\label{subsec:siam_hamiltonian}
Strongly correlated materials in thermodynamic equilibrium are often described 
by the Fermi-Hubbard model~\cite{hubbard_1963}. In this model, electrons can hop 
between adjacent lattice sites with amplitude $t$, and lattice sites occupied by 
a pair of electrons experience an energy penalty $U$.
\begin{figure}[t!]
	\centering
	\subfloat[\label{fig:impurity_bath_sites}]{
		\includegraphics[width=.45\linewidth,valign=b]{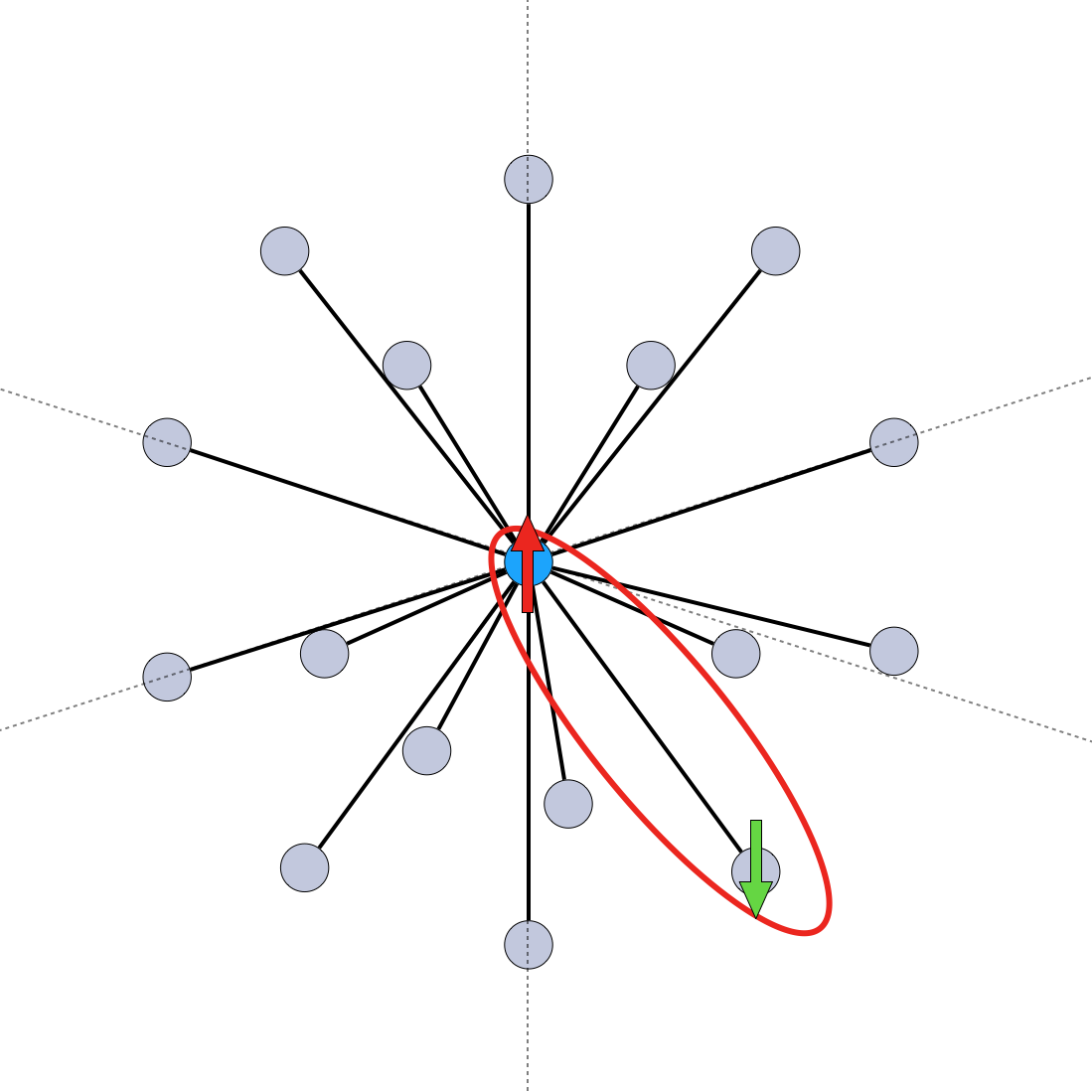}
	}	
	\hfill
	\subfloat[\label{fig:two_site_siam_half_filled_with_arrows}]{
		\includegraphics[width=.45\linewidth,valign=b]{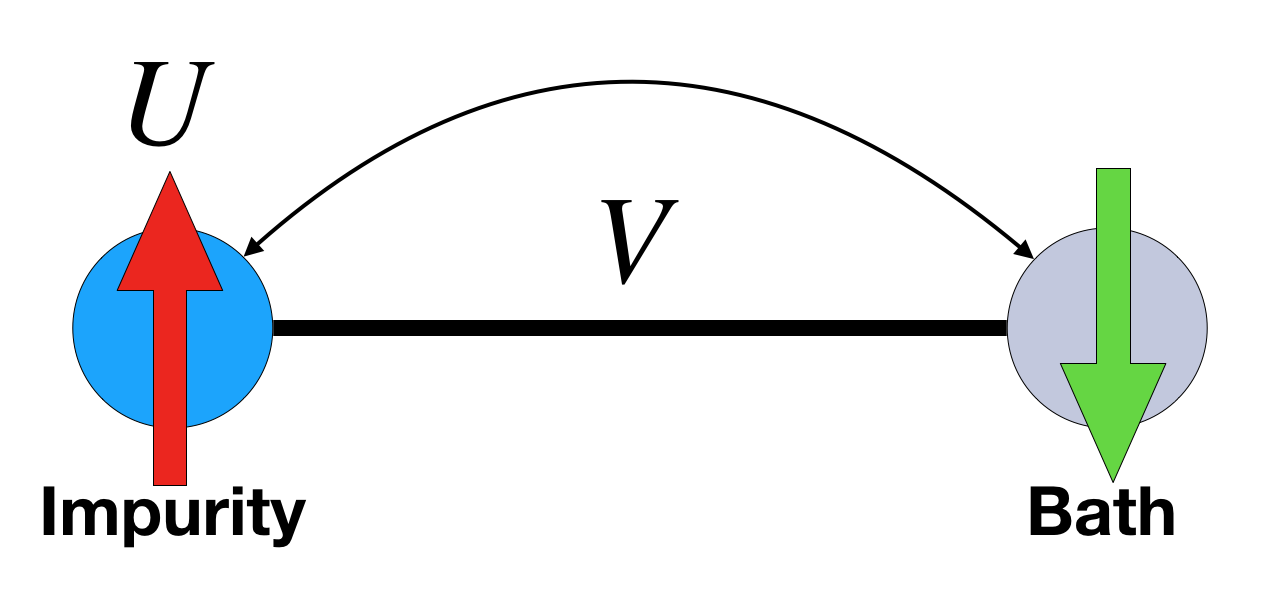}
	}
	\caption{(a) Hamiltonian-based DMFT approximates the many-body interactions 
		of strongly correlated systems with an impurity model. Electrons can occupy any 
		lattice site, but can only move between the impurity (blue) and a bath site 
		(grey). (b) In two-site DMFT, we use only a single bath site. In the half-filled 
		case, the system dynamics are now described by two parameters, $U$ and $V$, the 
		on-site interaction and hybridization parameter respectively.
		\label{fig:hamiltonian_based_dmft}}
\end{figure}

DMFT translates the many-body problem of the Hubbard model to a single-site 
impurity model. This reformulation is desirable because the problem then becomes amenable to various impurity solvers \cite{werner_comanac_demedici_troyer_millis_2006, rubtsov_savkin_lichtenstein_2005, haule_2007, werner_millis_2006}. To do this mapping, the interactions between the impurity and the surrounding fermions are represented as a time-varying mean-field, which the impurity site can exchange electrons with. The purpose of DMFT is to self-consistently determine a mean field such that the retarded impurity Green's function is equal to the local retarded lattice Green's function,
\begin{equation}
\label{eq: green_function_equality}
G^{R}_{\mathrm{imp}}(\omega) = G^{R}_{\mathrm{latt}, jj}(\omega).
\end{equation}
This mapping from a lattice model to an impurity model is exact in the limit 
where the number of spatial dimensions goes to infinity 
\cite{metzner_vollhardt_1989}. 
Here we consider the Fermi-Hubbard model embedded in an infinite dimensional Bethe lattice~\cite{georges_kotliar_krauth_rozenberg_1996}, as has been done previously for hybrid quantum DMFT~\cite{kreula_garcia-alvarez_lamata_clark_solano_jaksch_2016, rungger_fitzpatrick_chen_alderete_apel_2019}. To account for a lattice model with infinite coordination number $z \rightarrow \infty$, the Hubbard hopping amplitude $t$ needs to scale as $t \sim t^* / \sqrt{z}$ to avoid a diverging kinetic energy per lattice site. This defines a new constant, $t^*$, which is the Hubbard hopping amplitude in infinite dimensions.

In Hamiltonian-based DMFT, the mean-field is parametrised by a set of 
non-interacting bath sites, as shown in Fig.~\ref{fig:impurity_bath_sites}. 
This formulation of the impurity model is particularly conducive to being solved using a quantum computer, as for a given Hamiltonian $\hat{H}$, it requires evaluating the time evolution operator ${\hat{U}(\tau)} = \exp({-i\hat{H}\tau/\hbar})$. This is known to be exponentially faster on a quantum computer \cite{lloyd_1996}.

The self-consistency condition in Eq.~(\ref{eq: green_function_equality}) can only be satisfied exactly for an infinite number of bath sites. Here we consider the minimal implementation of Hamiltonian-based DMFT which involves just two sites - one for the impurity and another to approximate the mean field, see Fig.~\ref{fig:two_site_siam_half_filled_with_arrows}. This model is known as two-site DMFT~\cite{potthoff_2001}, and provides an approximate yet qualitatively correct description of strongly correlated phenomena in the Hubbard model. The system is now described by the SIAM Hamiltonian
\begin{align} 
\label{eq: two_site_general_hamiltonian}
\hat{H}_{\mathrm{SIAM}} = & U \hat{n}_{1\downarrow} \hat{n}_{1\uparrow} 
- \mu\sum_{\sigma}\hat{n}_{1\sigma} + 
\sum_{\sigma}\epsilon_{c}\hat{c}^{\dagger}_{2\sigma}\hat{c}_{2\sigma}
\notag \\
&  + \sum_{\sigma}V(\hat{c}^{\dagger}_{1\sigma}\hat{c}_{2\sigma} + \mathrm{H.c.}),
\end{align}
where $\hat{c}_{j,\sigma}^{\dagger}$ ($\hat{c}_{j,\sigma}$) is the fermionic 
creation (annihilation) operator, $\hat{n}_{j,\sigma} = \hat{c}_{j,\sigma}^{\dagger}\hat{c}_{j,\sigma}$ is the number operator acting on site $j$ with spin component $\sigma \in \set{\uparrow,\downarrow}$, $U$ is the same on-site interaction as in our original lattice model and $\mu$ is the impurity chemical potential. 

In general, the bath site energy  $\epsilon_{c}$ and the hybridization between the two sites $V$ need to be determined such that the self-consistency condition in Eq.~(\ref{eq: green_function_equality}) is approximately satisfied. In the following we focus on the half-filled case, which exhibits interesting effects such as the metal-insulator transition~\cite{lange_1998} and maximal antiferromagnetic spin correlations~\cite{cheuk_nichols_lawrence_okan_zhang_khatami_2016}. In this case, $\mu = {U}/{2}$ and $\epsilon_{c} = 0$, such that 
Eq.~(\ref{eq: two_site_general_hamiltonian}) reduces to
\begin{equation}
\label{eq: two_site_half_filled_hamiltonian}
\hat{H}_{\mathrm{SIAM}} = U \hat{n}_{1\downarrow} \hat{n}_{1\uparrow} - 
\frac{U}{2}\sum_{\sigma}\hat{n}_{1\sigma} + 
\sum_{\sigma}V(\hat{c}^{\dagger}_{1\sigma}\hat{c}_{2\sigma} + \mathrm{H.c.}).
\end{equation}
The hybridization parameter $V$ in Eq.~(\ref{eq: two_site_half_filled_hamiltonian}) is now the only free parameter 
that needs to be determined for a given $U$ such that Eq.~(\ref{eq: green_function_equality}) is approximately fulfilled. 
This self-consistency condition is shown to be equivalent to satisfying~\cite{potthoff_2001}
\begin{equation}
\label{eq: self_consistency_two_site_half_filled}
V^2 = Z{t^{*}}^{2}, 
\end{equation}
where $Z$ is the quasiparticle weight, which physically represents both the sign and magnitude of interactions in a Fermi liquid \cite{doggen_kinnunen_2015}.

Determining $V$ can be achieved via an iterative procedure incorporating a quantum processor and classical feedback loop, which
we describe in the next Sec.~\ref{subsec:quantum_circuits}.

\subsection{Hybrid quantum-classical DMFT routine \label{subsec:quantum_circuits}}
\begin{figure}[t!]
	\centering
	\includegraphics[width=\linewidth]{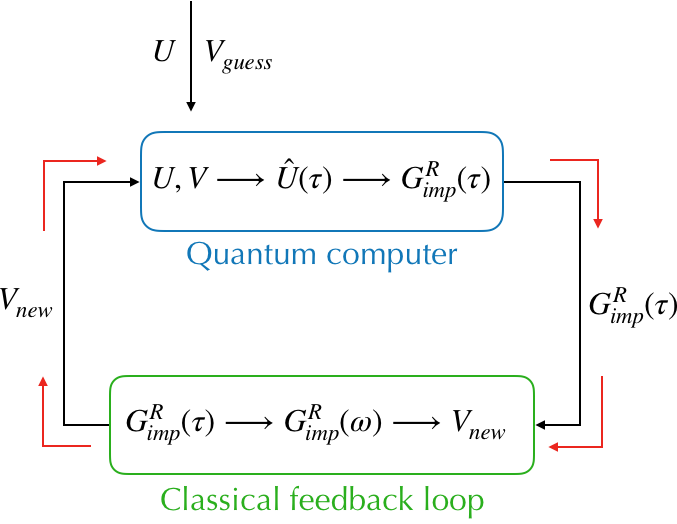}
	\caption{Diagram of hybrid quantum-classical DMFT. 
		For given on-site interaction energy $U$, we iteratively discover the
		hybridization parameter $V$ such that Eq.~(\ref{eq: self_consistency_two_site_half_filled})
		is satisfied. 
		For the first iteration, we 
		start with a guess $V_{\mathrm{guess}}$. We use a quantum computer to compute 
		the impurity Green's function $G^{R}_{\mathrm{imp}}(\tau)$, followed by a 
		classical optimiser to suggest an improved hybridization $V_{\mathrm{new}}$. The 
		full loop is iterated until self-consistency is reached, such that $V_{\mathrm{new}} = V$.}
	\label{fig:dmft_routine_with_arrows}
\end{figure}

The iterative process of hybrid quantum-classical DMFT  is illustrated in 
Fig.~\ref{fig:dmft_routine_with_arrows} and consists of the following 
steps~\cite{kreula_garcia-alvarez_lamata_clark_solano_jaksch_2016}:
\begin{enumerate}[wide, labelwidth=!, labelindent=0pt]
	\item Set the value of the impurity on-site interaction energy $U$.
	\item Make an initial guess for the value of the hybridization parameter $V$.
	\item Obtain the impurity Green's function $\imag G^{R}_{\mathrm{imp}}(\tau)$ 
	from the quantum computer as a function of time $\tau$ ($\imag$ is the imaginary unit). 
	\label{item:calculate_gf}
	\item At half-filling the impurity Green's function has the form
	\begin{equation}
	\imag G^{R}_{\mathrm{imp}}(\tau) = \alpha \cos(\omega_{1}\tau) + (1-\alpha) 
	\cos(\omega_{2}\tau).
	\end{equation} 
	Using the result for $\imag G^{R}_{\mathrm{imp}}(\tau)$ obtained from the quantum computer, find the 
	best fit for the parameters $\alpha, \omega_{1}$ and $\omega_{2}$, which make up 
	the residues and poles of $G^{R}_{\mathrm{imp}}(\omega)$ respectively. In contrast to other works~\cite{kreula_garcia-alvarez_lamata_clark_solano_jaksch_2016, keen_maier_johnston_lougovski_2020}, we use the normalization iG(0)=1 to reduce the number of fitting parameters from four to three.
	\item Calculate the quasiparticle weight according to 
	\begin{equation}
	\label{eq:z_formula_karsten_notes_16}
	Z = \left[V^4 \left(\frac{\alpha}{\omega_{1}^4} + 
	\frac{1-\alpha}{\omega_{2}^4}\right)\right]^{-1}.
	\end{equation}
	If the values for $Z$ and $V$ satisfy Eq. (\ref{eq: self_consistency_two_site_half_filled}), then self-consistency has been reached.
	\item Otherwise, update the hybridization parameter $V$ to one that would be self-consistent with the current system (i.e., $V_{\mathrm{new}} = \sqrt{Z}t^*$) and repeat from step 3.
	
\end{enumerate}

Next we show how the impurity Green's function can be measured using a 
quantum computer as required in step~\ref{item:calculate_gf}. To do this, we first map the impurity model onto a qubit system. Applying a Jordan-Wigner transformation \cite{jordan_wigner_1928} to Eq.~(\ref{eq: two_site_half_filled_hamiltonian}), we obtain
\begin{equation}
\label{eq: jw_transformed_siam}
\hat{H}_{\mathrm{SIAM}} = \frac{U}{4}(\hat{\sigma}_{1}^{z}\hat{\sigma}_{3}^{z}) 
+ \frac{V}{2}(\hat{\sigma}_{1}^{x}\hat{\sigma}_{2}^{x} + 
\hat{\sigma}_{1}^{y}\hat{\sigma}_{2}^{y} + 
\hat{\sigma}_{3}^{x}\hat{\sigma}_{4}^{x} + 
\hat{\sigma}_{3}^{y}\hat{\sigma}_{4}^{y}),
\end{equation}
where $\hat{\sigma}_{n}^{\alpha}$ is the Pauli operator $\alpha \in \set{x,y,z}$ acting on qubit $n$. As part of this process, we assign two qubits to represent each electronic site, due to its occupation and spin degrees of freedom.

\begin{figure}[t!] 
	\centering 
	\includegraphics[width=\linewidth]{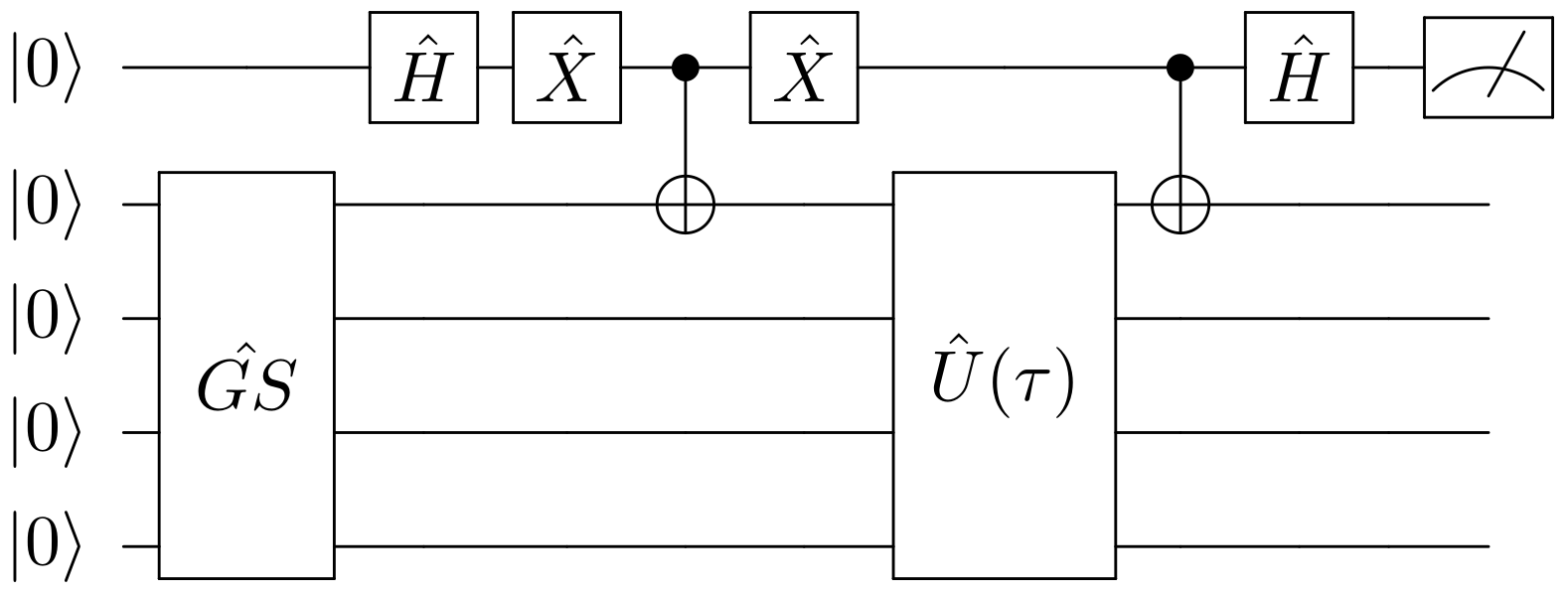}
	\caption{Quantum circuit used to calculate the expectation value $\langle 
		\hat{\sigma}_{1}^{x} {\hat{U}^{\dagger}(\tau) \hat{\sigma}_{1}^{x} 
			\hat{U}(\tau)}\rangle$. The work qubits are first prepared into the ground state 
		of the SIAM Hamiltonian using the sub-circuit $\hat{GS}$. They are then acted on 
		by entangling gates with the ancilla qubit and the time evolution operator 
		$\hat{U}(\tau)$. The ancilla qubit itself undergoes single-qubit Hadamard 
		$\hat{H}$ and bit-flip $\hat{X}$ gates. Repeated measurements of the ancilla in 
		the $\hat{\sigma}_{z}$, $\hat{\sigma}_{y}$ bases build up the real and imaginary 
		parts of the expectation value respectively.}
	\label{fig:scattering_symbolic}
\end{figure}

Next, we note that the impurity Green's function can be written as~\cite{kreula_garcia-alvarez_lamata_clark_solano_jaksch_2016}

\begin{equation}
\label{eq: gf}
G^{R}_{\mathrm{imp}}(\tau) = \theta(\tau)[G^{>}_{\mathrm{imp}}(\tau) - 
G^{<}_{\mathrm{imp}}(\tau)],
\end{equation}

where $\theta$ is the heavyside step function and the greater and lesser Green's functions are defined as

\begin{align}
G^{>}_{\mathrm{imp}}(\tau) &= -\imag \langle \hat{c}_{1\sigma}(\tau) 
\hat{c}^{\dagger}_{1\sigma}(0)\rangle , \\
G^{<}_{\mathrm{imp}}(\tau) &= \imag \langle \hat{c}^{\dagger}_{1\sigma}(0) 
\hat{c}_{1\sigma}(\tau) \rangle, 
\end{align}

respectively, where the average is computed in the ground-state of Eq.~(\ref{eq: two_site_half_filled_hamiltonian}). We apply a Jordan-Wigner transformation again, this time to Eq.~(\ref{eq: gf}), to express the impurity Green's function as
\begin{equation}
\label{eq: Gimp}
\imag G^{R}_{\mathrm{imp}}(\tau) = \mathrm{Re}[\langle \hat{\sigma}_{1}^{x} 
{\hat{U}^{\dagger}(\tau) \hat{\sigma}_{1}^{x} \hat{U}(\tau)}\rangle],
\end{equation}
where  
\begin{align}
\hat{U}(\tau) = 
\exp({-\imag\hat{H}_{\mathrm{SIAM}}\tau/\hbar})
\end{align}
is the time evolution operator. 
We evaluate $\imag G^{R}_{\mathrm{imp}}(\tau)$ via the quantum circuit shown in Fig.~\ref{fig:scattering_symbolic}, which we call the Green's function circuit. Based 
on the findings of~\cite{paz_roncaglia_2003}, we construct the expectation value 
$\mathrm{Re}[\langle \hat{\sigma}_{1}^{x} {\hat{U}^{\dagger}(\tau) 
	\hat{\sigma}_{1}^{x} \hat{U}(\tau)}\rangle]$ through repeated measurements of 
the ancilla qubit in the $\hat{\sigma}_{z}$ basis.  Notably, this circuit 
requires measuring only one qubit, which is true even as we increase the number 
of bath sites in the impurity model. 

In order to represent the time-evolution operator $\hat{U}$  in Eq.~(\ref{eq: Gimp}) 
in terms of quantum logic gates, 
we approximate it by a first order Suzuki-Trotter  decomposition~\cite{suzuki_1976} as shown 
in~\cite{kreula_garcia-alvarez_lamata_clark_solano_jaksch_2016}. 
By executing the Green's function circuit 
several times with different numbers of Trotter steps, we numerically 
reconstruct $\imag G^{R}_{\mathrm{imp}}$ as a function of~$\tau$. 
The circuit $\hat{GS}$ in Fig.~\ref{fig:scattering_symbolic} which prepares 
the ground state of the SIAM Hamiltonian can be obtained via arbitary state 
preparation techniques~\cite{shende_bullock_markov_2006}. 
\subsection{Noise model}\label{subsec:noise_models}
Next we describe the noise model that we use in our simulations of NISQ devices presented in Sec.~\ref{sec:results}. Our model, implemented using Qiskit, accounts for both imperfections in qubits and gates. It is applied to all operations allowed in our emulator, made up of the $\hat{U}_{1}$, $\hat{U}_{2}$ and $\hat{U}_{3}$ single-qubit gates (see Appendix \ref{appendix:single_qubit_gates}), the CNOT two-qubit gate and measurement.

Firstly, when an operation is applied to a qubit, we model
the qubit to undergo thermal relaxation 
based on its lifetimes $\tau_{1}$, $\tau_{2}$ and the gate time $t$, 
where $\tau_{1}$ and $\tau_{2}$ are the relaxation and dephasing time constants respectively. For simplicity, we set $\tau_{1} = \tau_{2} = \tau$ in this work and estimate operation times using guidance from both the literature 
\cite{baekkegaard_kristensen_loft_andersen_petrosyan_zinner_2019, 
linke_maslov_roetteler_debnath_figgatt_landsman_wright_monroe_2017, 
kjaergaard_schwartz_2019} and example noise models given in Qiskit. To calculate the probability of thermal relaxation during a two-qubit gate, we tensor product the single-qubit error channels of each of the two qubits involved.

Secondly, we model the imperfections of quantum gates using a 
depolarizing quantum error channel~\cite{nielsen:2011:QCQ:1972505}. When applied to a single qubit, this has the form
\begin{equation}
\epsilon(\rho) = (1 - \lambda)\rho + \frac{\lambda}{3}\left(\sigma^{x} \rho 
\sigma^{x} + \sigma^{y}\rho \sigma^{y} + \sigma^{z}\rho \sigma^{z} \right),
\end{equation}
where $\rho$ is the density matrix of the qubit. The physical interpretation of this error channel is that when a gate is applied, an additional Pauli operation occurs with probability $\lambda$. The depolarizing channel is often used to characterise quantum noise~\cite{bennett_divincenzo_smolin_wootters_1996, cafaro_mancini_2010}, particularly as a worst case scenario where we have little information about the true noise channels, which makes it an apt description of NISQ devices. We subsequently implement the depolarizing channel for both single and two-qubit gates.
\begin{figure}[t!]
	\centering
	\includegraphics[width=\linewidth]{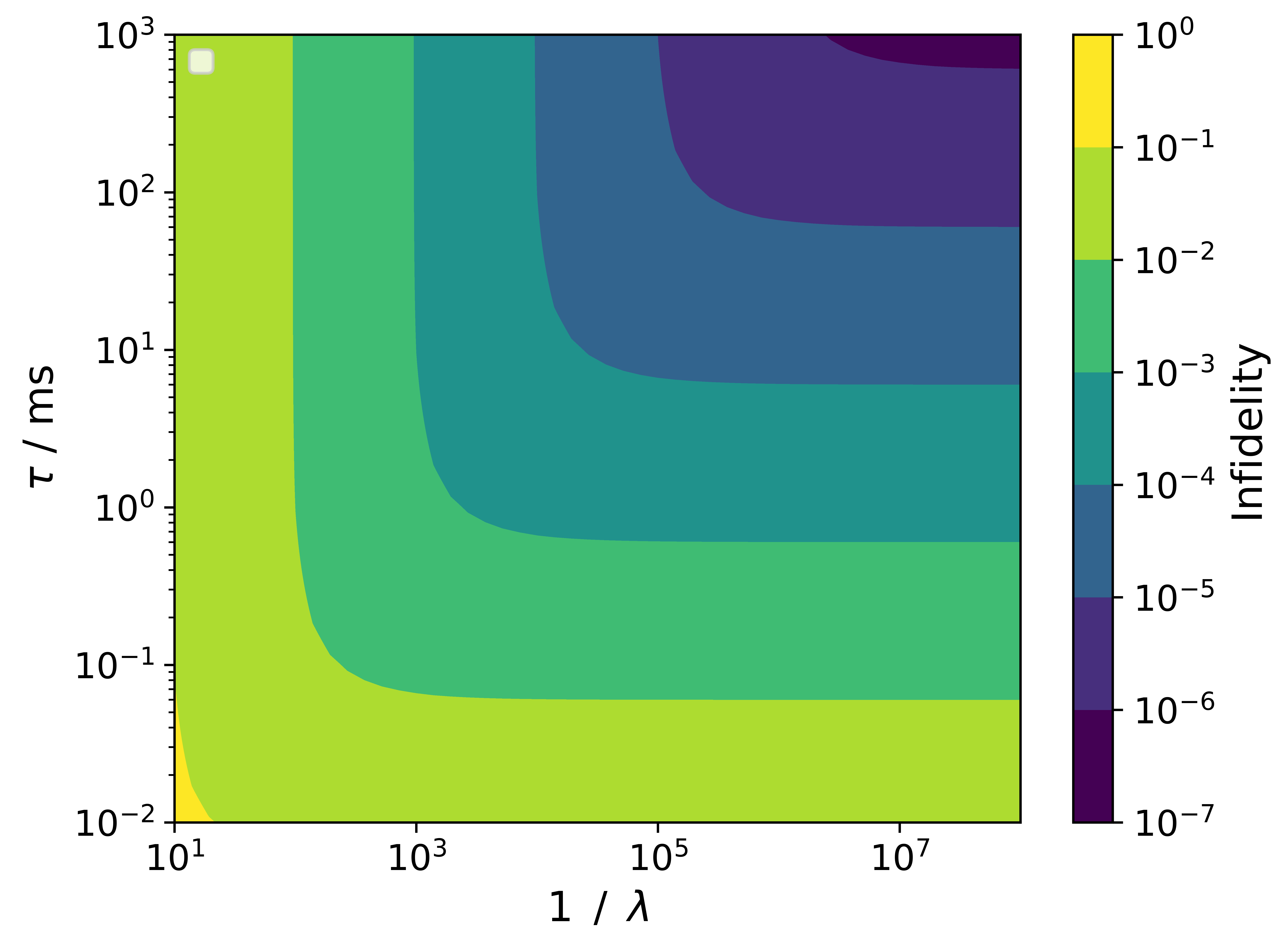}
	\caption{CNOT gate infidelity for different noise model parameters. The 
		qubit lifetime $\tau$ is used to calculate the probability of thermal relaxation 
		occurring. Additionally, a two-qubit depolarizing channel is applied with 
		probability parameter $\lambda$. For this particular noise model, we set the 
		CNOT gate time to be 300ns. Note that the color coding of the infidelity utilises a discretized logarithmic scale.}
	\label{fig:cx_fidelity_map}
\end{figure}

We combine the thermal relaxation and depolarizing error channels to produce a realistic emulation of noisy quantum computers \cite{knill_2005, Qiskit}. From this, individual fidelities can be extracted for any operation - including single-qubit gates, two-qubit gates and measurements. For example, Figure~\ref{fig:cx_fidelity_map} shows the infidelity of the CNOT gate, as a function of the noise model parameters. We see that if the depolarizing error is negligible, (i.e., small values of $\lambda$), the gate infidelity only depends on qubit lifetime. Conversely, in the limit of very long qubit lifetimes $\tau$, the depolarizing error becomes the dominant source of error. Moreover, we find that in this case the gate infidelity is equal to the value of the depolarizing parameter $\lambda$. It follows that achieving high fidelity requires a combination of both long qubit lifetime and low depolarizing error probability.

\section{Results}\label{sec:results}
We now implement the DMFT routine described in section \ref{sec:model} in Python, constructing the relevant quantum circuits in Qiskit. In section~\ref{subsec:fidelity_requirements_original}, we find the minimum number of Trotter steps required to reproduce the analytic two-site DMFT solution and consider the number of shots of the Green's function circuit required to mitigate statistical errors. We use these results to subsequently find the lowest gate fidelities that can produce accurate results compared to the noiseless solution. In section \ref{subsec:isl} we apply incremental structural learning (ISL), our circuit recompilation algorithm, and compare by how much the minimum hardware requirements change.
\subsection{Fidelity requirement of original scheme} \label{subsec:fidelity_requirements_original}
\begin{figure}[]
	\centering
	\includegraphics[width=\linewidth]{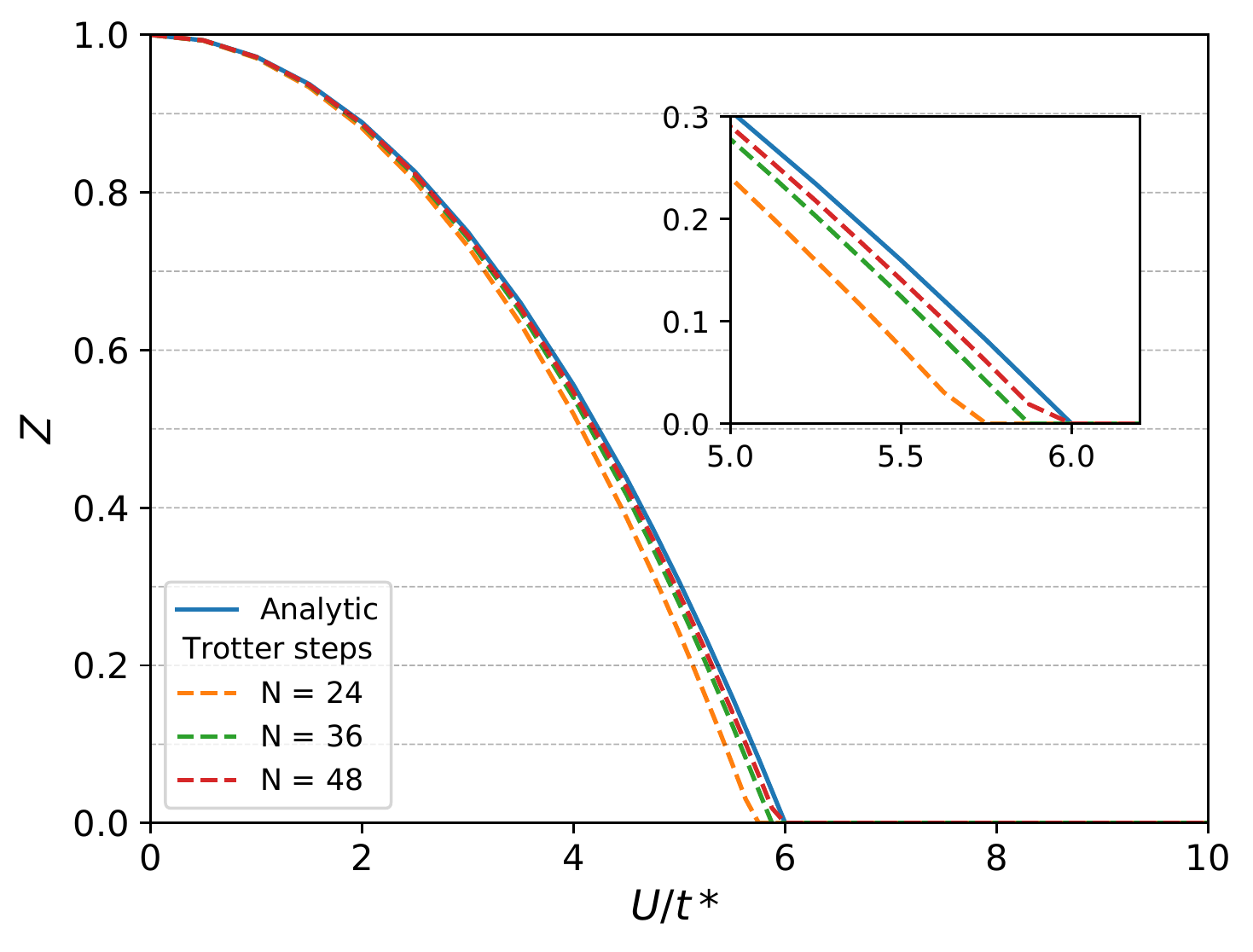}
	\caption{Quasiparticle weight $Z$ as a function of interaction strength $U$ and the Hubbard hopping amplitude in infinite dimensions $t^*$. For a given interaction strength, we iteratively obtain a self-consistent $Z$ for 24, 36 and 48 Trotter steps and compare against the analytic solution. The inset focuses on the region near the critical value $U_{c} = 6.0t^*$.\label{fig:zplot_statevector}}
	\label{fig:zplot_statevector_comparison_overview}
\end{figure}

We run the full DMFT scheme described in Sec.~\ref{sec:model} 
using a noiseless statevector simulator for 
different numbers of Trotter steps. The results, seen in Fig. 
\ref{fig:zplot_statevector}, show excellent agreement of the converged 
quasiparticle weight $Z$ to the analytic solution \cite{potthoff_2001}, 
particularly in the conducting phase at $U < 3.0t^*$. As we approach the 
metal-insulator phase transition at $U=6.0t^*$,our hybrid algorithm underestimates the quasiparticle weight and in the cases of $N=24$ and $N=36$ Trotter steps, incorrectly identifies where the transition occurs. This is an expected consequence of the approximations made during a Trotter decomposition and can be rectified by increasing the number of Trotter steps. Indeed, for $N=48$ Trotter steps we see excellent agreement to the analytic solution, even at the phase transition.

To minimise circuit depth, we now focus on the $N=24$ Trotter steps case, which still provides accurate results in the range $2.0t^* < U < 3.5t^*$. In order to apply our noise model to the simulated hardware, we must first switch from using a statevector simulator to a measurement-based one. In doing so, we add a source of error to our simulation in the form of shot noise (i.e., the number of measurements required to build up an accurate expectation value). Through experimentation, we find 75,000 shots to be sufficient for the statistical error to be less than the error generated by our noise models. This is well within the capabilities of NISQ devices.

We then look to apply our noise model to test the performance of DMFT. Using Fig. \ref{fig:cx_fidelity_map}, we find noise model parameters on the boundary of each infidelity contour such that they correspond to CNOT gate fidelities separated by an order of magnitude each, e.g., 99\%, 99.9\%, 99.99\% and so forth. These parameters, $\tau$, $\lambda$, subsequently determine the single-qubit gate and measurement fidelities, which are always larger than the CNOT fidelity. We implement these parameters in our simulations and run DMFT, transpiling all quantum circuits in Qiskit with the "heavy" optimisation option.

\begin{figure}[t!]
	\centering
	\includegraphics[width=\linewidth]{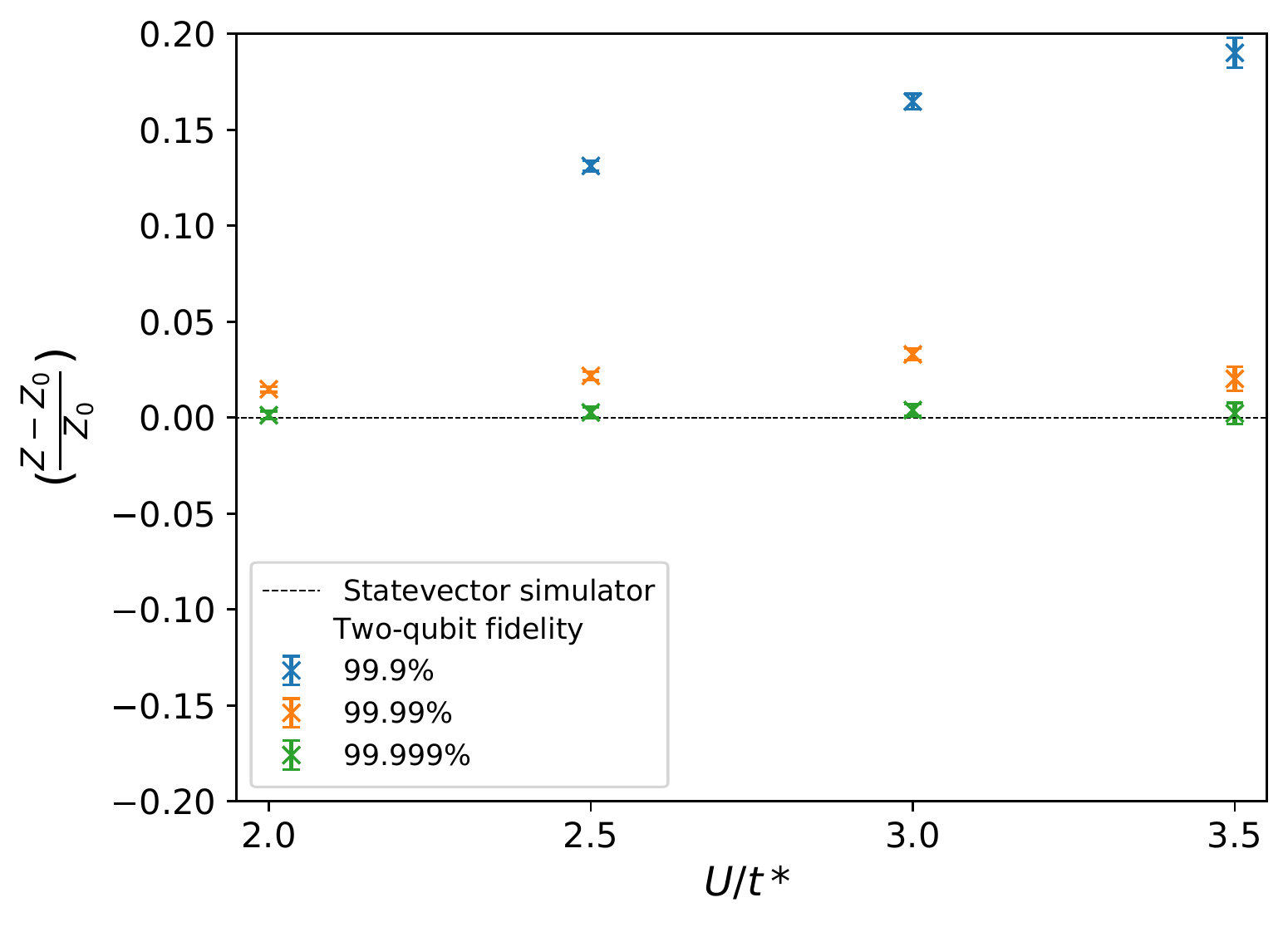}
	\caption{Relative quasiparticle weight as a function of on-site interaction strength $U$ for different two-qubit gate fidelities. Here $Z_{0}$ is the 24 Trotter step statevector simulator result shown in Fig. \ref{fig:zplot_statevector}. For definition of error  bars see text.}
	\label{fig:zplot_noisemodel_cnot}
\end{figure}

As shown in Fig. \ref{fig:zplot_noisemodel_cnot}, simulations with higher gate fidelities produce quasiparticle weights closer to the noiseless solution, as obtained on the statevector simulator. We observe that a two-qubit fidelity of 99.9\% is not sufficient for DMFT to converge consistently, as shown by the absence of a result in the $U=2.0t^*$ case. Increasing the fidelity to 99.99\% (noise parameters A) allows DMFT to converge to a quasiparticle weight within 4\% of the exact solution. The full details of these noise parameters and all others referenced in this section are shown in~Table~\ref{table:noise_models}.

The error bars shown in Fig.~\ref{fig:zplot_noisemodel_cnot} are determined as follows. We find that in the presence of noise, DMFT oscillates around the self-consistent solution without settling to a finite value. To account for this, after the self-consistent threshold is met \cite{self_consistent_threshold} we run 50 additional iterations and take the average quasiparticle weight as our solution. We then use the standard deviation $\sigma$ of these iterations to produce the error bars in Fig.~\ref{fig:zplot_noisemodel_cnot} of size $2\sigma$.

The low-fidelity results in Fig.~\ref{fig:zplot_noisemodel_cnot}, shown by the blue and orange data points, demonstrate that the magnitude of the error bars does not fully account for the deviation from the exact solution. We find that the applied noise channels dampen the oscillations of the impurity Green’s function, restricting its ability to represent the analytically correct solution. If the gate error is too large, and the impurity Green’s function too misshapen, the quasiparticle weight at each DMFT iteration will jump too much to converge. However, subsequent steps may fall within the self-consistency tolerance~\cite{self_consistent_threshold} at a larger incorrect solution $Z > Z_0$. This is because of the non-linear relationship between $V_\mathrm{new}$ and $V$, evident when substituting the self-consistency condition Eq.~(\ref{eq: self_consistency_two_site_half_filled}) into Eq.~(\ref{eq:z_formula_karsten_notes_16}). In particular, we find $V_{\mathrm{new}} \propto V^{-2}$ and thus large values of $V$ result in smaller step sizes, which explains the spurious convergence observed for the low-fidelity results in Fig.~\ref{fig:zplot_noisemodel_cnot}.

The high-fidelity result, with two-qubit gate fidelity of 99.999\% (noise parameters B) are shown by the green data points in Fig.~\ref{fig:zplot_noisemodel_cnot}. We find that this represents the maximum noise that can be tolerated whilst reproducing the statevector simulator within the noise-induced error bars.

\subsection{Circuit reduction using incremental structural 
	learning}\label{subsec:isl}

The total noise incurred in the execution of a quantum circuit scales 
exponentially with the number of logic gates. Therefore, we focus on lowering 
the fidelity requirements of quantum DMFT by reducing the length of the Green's 
function circuit. This is achieved using a circuit recompilation technique we 
call incremental structural learning (ISL). This follows many recent successes 
in using variational quantum algorithms to recompile quantum circuits, from 
which we draw inspiration \cite{jones_benjamin_2018, 
	khatri_larose_poremba_cincio_sornborger_coles_2019, ostaszewski_mateusz_grant_benedetti_2019, ostaszewski_mateusz_grant_benedetti_2019}.

We significantly reduce the required circuit depth in two ways. First, we use the variational quantum eigensolver \cite{peruzzo_mcclean_shadbolt_yung_zhou_love_aspuru-guzik_obrien_2014} to find an approximate representation of the circuit $\hat{GS}$ which prepares the ground state of the SIAM Hamiltonian. In this way, we reduce the depth of $\hat{GS}$ from 72 using exact initialisation technique to 4. Note that in this approach the ground state of the SIAM Hamiltonian does not need to be calculated on a classical computer. 

Second, we use ISL to reduce  the depth of the full Green's function circuit 
shown in Fig.~\ref{fig:scattering_symbolic}.
For an arbitrary quantum circuit $\hat{A}$, the goal of ISL is to find a 
shallower circuit $\hat{B}$ which has approximately the same action on an input 
state $\ket{\psi}$, such that $\hat{A}\ket{\psi}=\hat{B}\ket{\psi}$.  
The details of ISL are presented in Appendix~\ref{appendix:ISL_description}. 

\begin{figure}[t!]
	\centering
	\includegraphics[width=\linewidth]{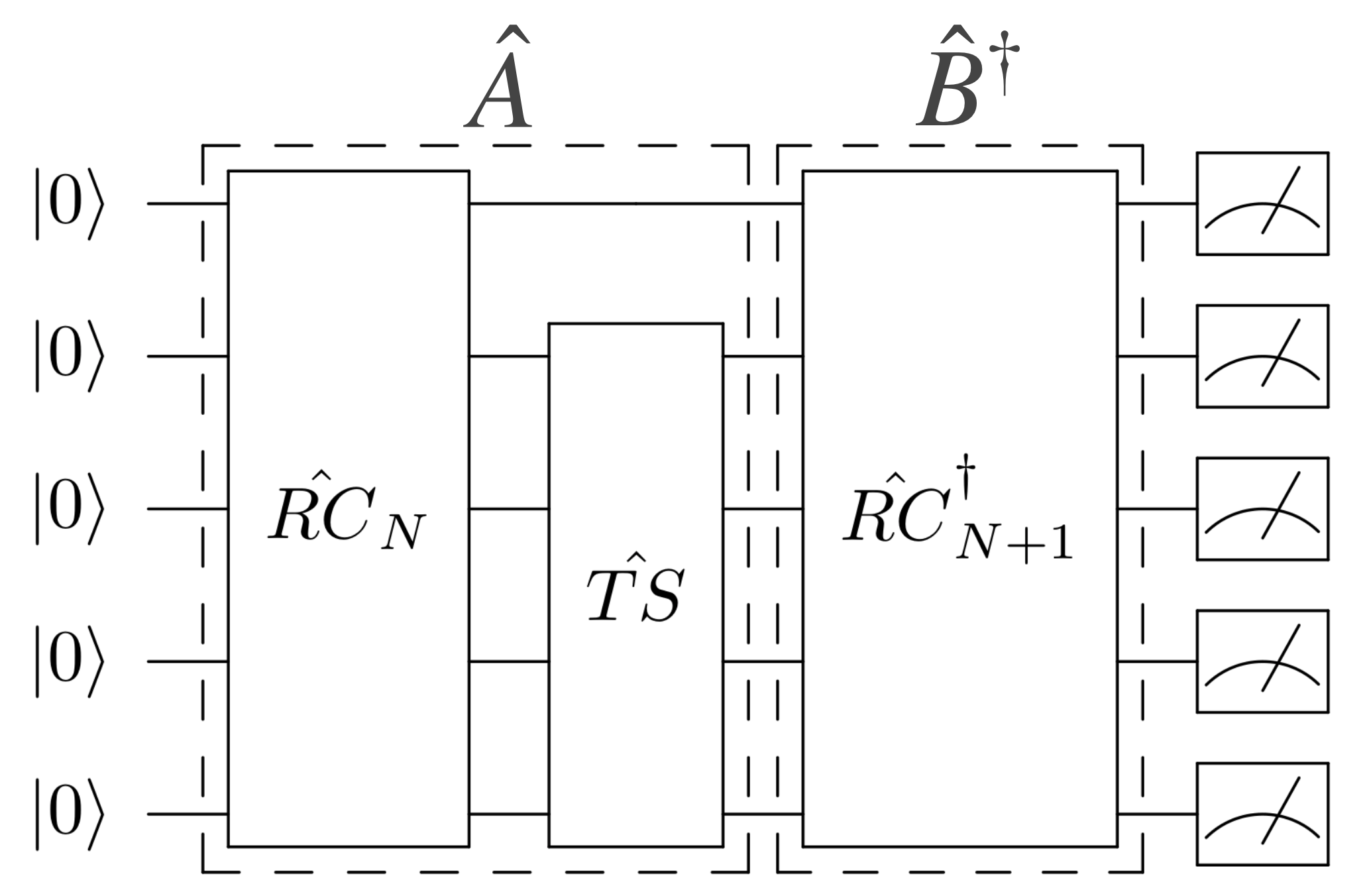}
	\caption{General structure of the ISL procedure. A recompilation target 
		$\hat{A}$ for the $N+1$ Trotter steps Green's function circuit is created by 
		adding one exact Trotter step to $\hat{RC}_{N}$, the recompiled circuit for $N$ 
		Trotter steps. A solution is constructed by trying to find a circuit $\hat{B^{\dagger}} = \hat{RC}^{\dagger}_{N+1}$ that approximately acts as the inverse of $\hat{A}$ but with fewer gates. The ansatz for $\hat{B^{\dagger}}$ is built up iteratively until the overlap between the output state and the input state $\ket{0}^{\otimes n}$ is sufficiently large. Details on the specifics of this procedure can be found in Appendix \ref{appendix:ISL_description}. 
	}
	\label{fig:isl_general_circuit_with_A}
\end{figure}

We apply ISL iteratively for every Trotter step as illustrated in 
Fig.~\ref{fig:isl_general_circuit_with_A}. Generally, for the $N+1$ Green's function circuit, where $N$ is the number of Trotter steps, we use the ISL solution of the previous $N$ Green's function circuit and add one exact Trotter step to create $\hat{A}$. 

By using this iterative approach, the depth of the ISL circuit does not scale with the number of Trotter steps being simulated. Therefore, the deepest circuit our algorithm needs to run is one exact Trotter step plus the two recompiled solutions, which has an average depth of 41. Once completed, ISL produces a Green's function circuit containing on average 6 two-qubit gates and 11 single-qubit gates for any number of Trotter steps. This is in contrast to the 24 Trotter step Green's function circuit in the original scheme, which contains 510 two-qubit gates and 752 single-qubit gates.

\begin{figure}[t!]
	\centering
	\includegraphics[width=\linewidth]{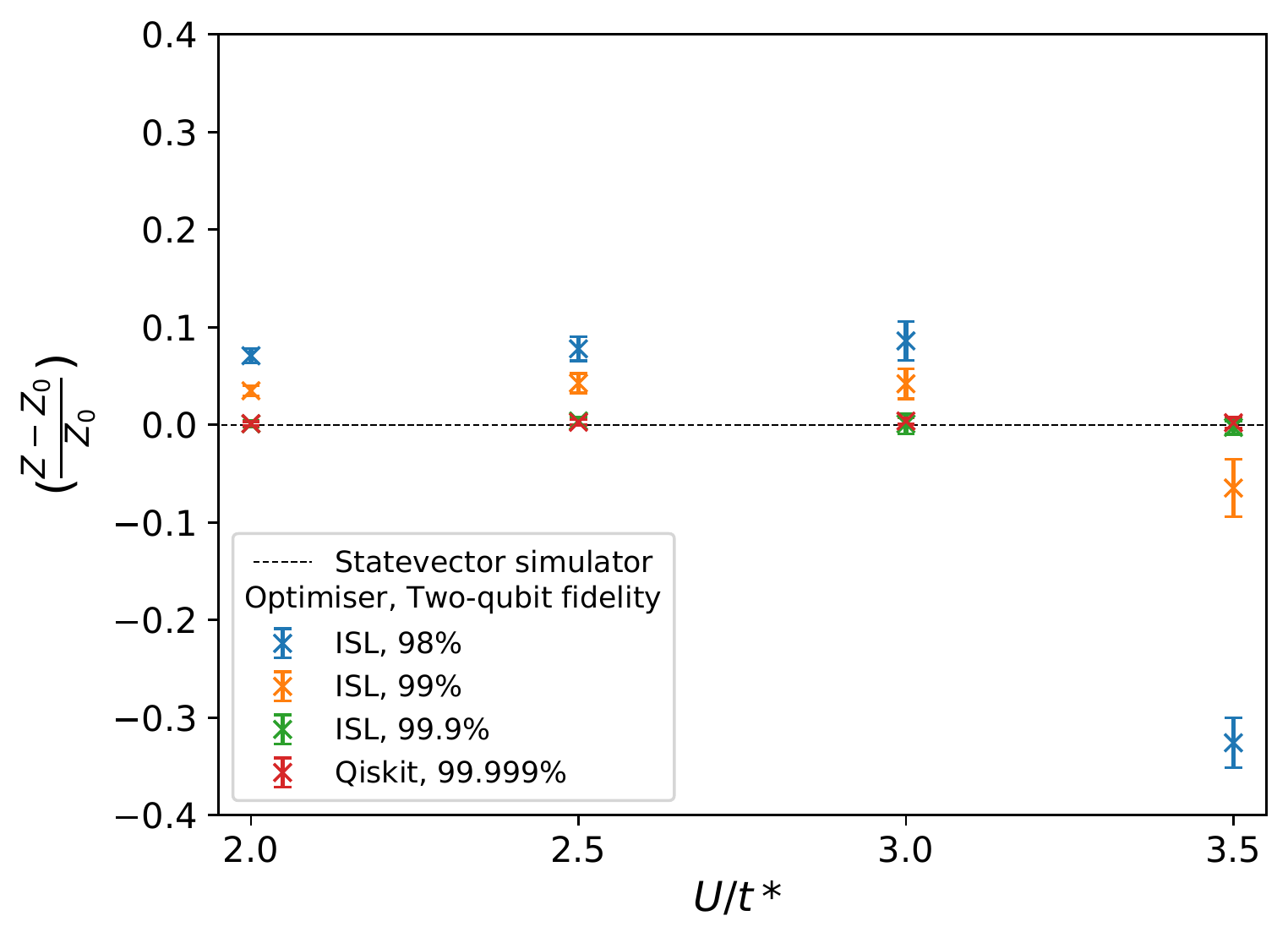}
	\caption{Relative quasiparticle weight as a function of on-site interaction strength $U$ for different two-qubit gate fidelities. Using incremental structural learning (ISL), a quantum computer with two-qubit gate fidelity of 99.9\% can produce convergent DMFT results with perfect agreement to the noiseless solution.}
	\label{fig:zplot_noisemodel_isl}
\end{figure}

We rerun hybrid quantum DMFT, using the same noise parameters as in section \ref{subsec:fidelity_requirements_original}, but this time applying ISL to each Green's function circuit. In Fig.~\ref{fig:zplot_noisemodel_isl}, we show that in this case, a two-qubit fidelity of 98\% (noise parameters C) or 99\% (noise parameters D) is enough for DMFT to converge within 35\% or 10\% of the exact solution respectively. Furthermore, we find that a quantum computer with 99.9\% two-qubit gate fidelity (noise parameters E) is sufficient to produce results that perfectly agree with the statevector simulator within noise-induced error margins. Therefore, by applying ISL, we see a factor of 100 improvement in the noise tolerance of two-site hybrid DMFT compared to using non-approximate circuit recompilation techniques.

\begin{table}[b!]
	\resizebox{\linewidth}{!}{%
		\begin{tabular}{|l|lllllll|}
			\hline
			Noise parameters & $\tau$/ms & $\lambda$ & $\mathcal{F}(U_{1})$ & 
			$\mathcal{F}(U_{2})$ & $\mathcal{F}(U_{3})$ & $\mathcal{F}(\mathrm{CNOT})$ & 
			$\mathcal{F}(\mathrm{Measurement})$ \\ \hline
			A & 10 & 4e-5 & 0.99997 & 0.99997 & 0.99996 & 0.9999 & 0.9999 \\
			B & 100 & 4e-6 & 0.999997 & 0.999997 & 0.999996 & 0.99999 & 0.99999 \\
			C & 0.04 & 5e-3 & 0.996 & 0.995 & 0.994 & 0.980 & 0.975 \\
			D & 100 & 4e-6 & 0.997 & 0.997& 0.996 & 0.990 & 0.991 \\
			E & 1.1 & 4e-4 & 0.9997 & 0.99974 & 0.9996 & 0.999 & 0.999 \\ \hline
		\end{tabular}%
	}
	\caption{Parameters used to emulate different NISQ devices with our noise model. For a given qubit lifetime $\tau$ and depolarizing channel probability $\lambda$, the corresponding gate fidelities $\mathcal{F}$ can be obtained. The single-qubit gates, $\hat{U}_{1}$, $\hat{U}_{2}$ and $\hat{U}_{3}$ are defined in Appendix \ref{appendix:single_qubit_gates}.}
	\label{table:noise_models}
\end{table}

\section{Conclusion} \label{sec:conclusion}
In this work, we find that a previously proposed algorithm for hybrid quantum-classical DMFT can be accurately solved within noise-induced error margins, provided quantum hardware capable of executing 75,000 shots, two-qubit gate fidelity of 99.999\% and average single-qubit gate fidelity of 99.9997\% (noise parameters B). However, by finding shallow approximations of the Green's function circuits using our ISL recompiler, we show that DMFT can be self-consistently solved by quantum hardware with two-qubit and average single-qubit fidelities of 99\% and 99.7\% respectively (noise parameters D), within 10\% of the exact solution. These results are consistent with those for the implementation of the alternative DMFT scheme in~\cite{rungger_fitzpatrick_chen_alderete_apel_2019}, where a solution at U=4 was found within 2.6\% of the exact solution on an IBM device using circuit recompilation and SPAM error mitigation methods.  Note that the algorithm in~\cite{rungger_fitzpatrick_chen_alderete_apel_2019} requires one to calculate all excited states via the variational quantum eigensolver method, and thus it scales exponentially with the number of sites. Furthermore, we find that increasing the two-qubit and single-qubit fidelities to 99.9\% and 99.97\% respectively (noise parameters E) allows one to produce results in perfect agreement with the exact solution, within the noise-induced error bounds.

Excitingly, these findings show that our scheme could produce accurate results on noisy quantum computers in the near future. For superconducting qubit architectures, Google's \textit{Sycamore} 53 qubit device has two-qubit and single-qubit gate fidelities of 99.64\% and 99.85\% respectively \cite{arute_arya_babbush_bacon_bardin_2019}. Given that the total noise scales with the number of qubits, these figures suggest that our fidelity requirements could already be met by a smaller, high fidelity, 5 qubit device. 

A different perspective can be gained on the capabilities of NISQ computers if we consider quantum volume instead \cite{cross_bishop_sheldon_nation_gambetta_2019}. Using randomized circuit 
benchmarking, we calculate the quantum volume corresponding to noise parameters E to be 32. By comparison, IBM's recently announced 28 qubit \textit{Raleigh} device has the largest measured quantum volume to date of also 32. Given access to this device is planned for 2020, we expect our scheme to be runnable on real quantum hardware by the end of the year.

Looking forward, an open problem remains to determine the fidelity requirements for hybrid quantum-classical DMFT with more than just two sites. This is particularly true for achieving a quantum advantage, which would require more than 25 bath sites (50 qubits). Whilst the scalability of variational algorithms such as VQE and ISL is an open question, the number of gates in our scheme grows sub-exponentially with the number of DMFT sites. In this way, hybrid quantum-classical DMFT may prove to be another candidate for 
displaying quantum advantage before the era of fault-tolerant qubits.

\begin{acknowledgments}
	We acknowledge support from the EPSRC National Quantum Technology Hub in Networked Quantum Information Technology (EP/M013243/1) and the EPSRC Hub in Quantum Computing and Simulation (EP/T001062/1). MK and DJ acknowledge financial support from the National Research Foundation, Prime Ministers Office, Singapore, and the Ministry of Education, Singapore, under the Research Centres of Excellence program.
\end{acknowledgments}

\appendix

\section{Recompiling quantum circuits using incremental structural learning} 
\label{appendix:ISL_description}

ISL represents a special case of quantum circuit compilation, whereby the input state of the target circuit is always $\ket{\psi_{0}} = \ket{0}^{\otimes n}$. Since we wish to find an ansatz $\hat{B}^{\dagger}$, which acts as the inverse of a target $\hat{A}$, ISL minimises the cost function
\begin{equation}\label{eq:isl_cost_function}
C = 1 - \left|\bra{\psi_{0}}\hat{B}^{\dagger}\hat{A}\ket{\psi_{0}}\right|^2,
\end{equation}
where $\bra{\psi_{0}}\hat{B}^{\dagger}\hat{A}\ket{\psi_{0}}$ is the overlap 
between the input and output states of Fig. 
\ref{fig:isl_general_circuit_with_A}.

Instead of using a fixed ansatz for $\hat{B}^{\dagger}$, we incrementally build 
its structure layer-by-layer, evaluating the cost function each time. This 
approach offers the most flexibility to find the optimal solution, at the 
expense of greater computational cost. Nevertheless, structural ansatzes have 
seen notable success in hybrid algorithms such as ADAPT-VQE 
\cite{grimsley_economou_barnes_mayhall_2019}.
\subsection{Constructing $\boldmath\hat{B}^{\dagger}$}

\begin{figure}[t!]
	\centering
	\includegraphics[width=0.7\linewidth]{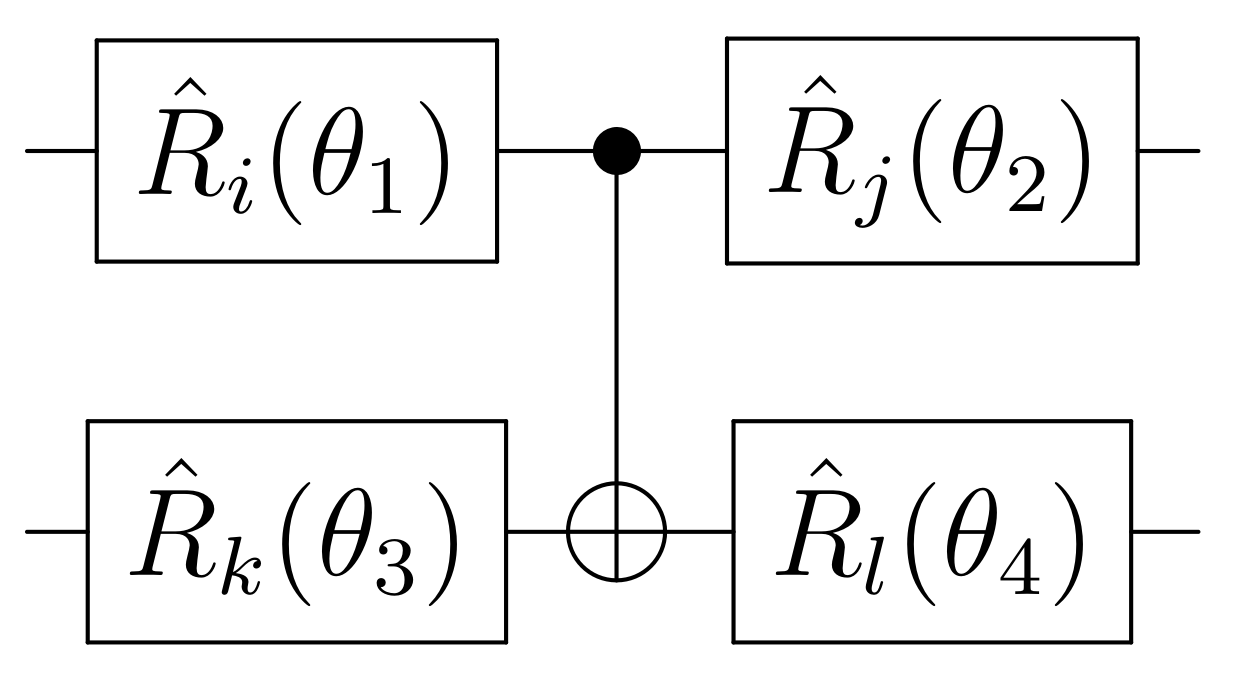}
	\caption{A thinly-dressed CNOT gate is a CNOT gate surrounded by 4 
		single-qubit rotation gates $\hat{R}_i(\theta)$, where $i \in \set{x,y,z}$ is 
		the axis of rotation and $\theta$ is the angle.}
	\label{fig:dressed_cnot}
\end{figure}

The ansatz $\hat{B}^{\dagger} = \hat{B}^{\dagger}_{n}...\hat{B}^{\dagger}_{1}$ consists of $n$  layers of $\hat{B}^{\dagger}_{i}$, where $\hat{B}^{\dagger}_{i}$ is a thinly-dressed CNOT 
gate as shown in Fig. \ref{fig:dressed_cnot}. We describe this as thinly 
dressed because the single-qubit gate rotations are restricted to one axis -  in 
contrast to the regular dressed CNOT gates in 
\cite{sharma_khatri_cerezo_coles_2019}. When adding the $i^{th}$ layer $\hat{B}^{\dagger}_{i}$, we must first decide which qubits 
should be acted on. To do this we evaluate the entanglement of formation $E$ 
\cite{wootters_1998} between each pair of qubits, which are in the state $\hat{B}^{\dagger}_{i-1}...\hat{B}^{\dagger}_{1}\hat{A}\ket{\psi_{0}}$. Practically, this is achieved by performing a partial trace over all other qubits and then calculating $E$ from the resulting mixed, bipartite state. We subsequently choose the qubit pair with the highest $E$ as the control and target qubits for the thinly-dressed CNOT gate of this layer $\hat{B}^{\dagger}_{i}$.

It is also possible that all qubit pairs have $E=0$. For example, the maximally 
entangled state $\ket {GHZ}$ does not have any pairwise local entanglement and 
will result in $E=0$ for all qubit pairs. In this case, we measure the expectation value $\langle\hat{\sigma}_{z}\rangle$ of each qubit. Since $\langle\hat{\sigma}_{z}\rangle = 0$ for the input qubits, we apply a thinly-dressed CNOT layer to the two qubits with the highest and second highest expectation values. 

One constraint that we impose on the choice of the control and target is that it 
must not be the same as the control and target for the previous layer. This is 
because in general, adding layers to different choices of control and target 
qubits allows us to explore a greater region of the available Hilbert space. This 
also avoids creating circuits with large depth but small numbers of gates. Hence, 
if the qubit pair with the highest $E$ is the same as in the previous 
layer, we choose different qubits with the two largest expectation values instead.

Once we have chosen the control and target qubits, we add the layer to 
$\hat{B^{\dagger}}$ with initial rotations $\theta=0$ about the z axis.

\subsection{Optimising}
After a layer is added, the axes and angles of rotation of the single-qubit 
gates are optimised using the \texttt{rotoselect} structural learning procedure 
\cite{ostaszewski_mateusz_grant_benedetti_2019}, with respect to minimising Eq. 
(\ref{eq:isl_cost_function}). This procedure works by fixing three of the gates 
and varying the rotation axes and angle for the remaining one. This is then repeated, sequentially cycling over the 4 rotation gates until a termination criterion is reached. We define this as when the reductions in the cost function between cycles is less than 1\%.

Once the single-qubit gates of this particular layer have been optimised, we 
then optimise the whole ansatz $\hat{B^{\dagger}}$ using \texttt{rotosolve} 
\cite{ostaszewski_mateusz_grant_benedetti_2019}. This procedure is similar to 
\texttt{rotoselect}, but doesn't involve optimizing the rotation gate axes.
\subsection{Terminating}
Once the \texttt{rotosolve} procedure is terminated, we perform standard non-approximate transpilation of $\hat{B^{\dagger}}$. Examples of this include the removal of both duplicate gates and rotation gates with very small angles. 

After this we take one final measurement of the cost function. If it is above a certain minimum threshold, we repeat the process again and add a new layer. If it is below the threshold, we terminate ISL and recursively invert all of the gates in the ansatz to return $\hat{B}$. Specifically for hybrid quantum-classical DMFT, we append the final ancilla operations to $\hat{B}$ and create a Green's function circuit.

\section{Definition of single-qubit gates} \label{appendix:single_qubit_gates}
The single-qubit unitary gates, $U_{1}$, $U_{2}$ and $U_{3}$, are defined 
as~\cite{Qiskit}

\begin{align}
\hat{U}_{3}(\theta, \phi, \lambda) & = \begin{pmatrix} \cos(\theta/2) & 
-e^{i\lambda}\sin(\theta/2) \\ e^{i\phi}\sin(\theta/2) & 
e^{i\lambda+i\phi}\cos(\theta/2) \end{pmatrix}\,, \\
\hat{U}_{2}(\phi, \lambda) & = \hat{U}_{3}(\pi/2, \phi, \lambda) = 
\frac{1}{\sqrt{2}} \begin{pmatrix} 1 & -e^{i\lambda} \\ e^{i\phi} & e^{i(\phi + 
	\lambda)} \end{pmatrix}\,,\\
\hat{U}_{1}(\lambda) & = \hat{U}_{3}(0, 0, \lambda) = \begin{pmatrix} 1 & 0 \\ 
0 & e^{i \lambda} \end{pmatrix}\,.
\end{align}

Although the $\hat{U}_{3}$ gate is universal, it is useful to distinguish these three separate gates operations for noise modelling purposes. This is because the $\hat{U}_{1}$, $\hat{U}_{2}$ and $\hat{U}_{3}$ gates require 0, 1 and 2 X90 pulses respectively. This in turn affects how long it takes to run each gate.

\bibliography{bibliography.bib}

\end{document}